\documentclass[preprint,showpacs,showkeys,aps,prb]{revtex4}

\usepackage[dvips]{graphicx}

\begin{document}


\title{\bf Maxwell and Cattaneo's Time-Delay Ideas \\
Applied to Shockwaves and the Rayleigh-B\'enard Problem}  

\author{
Francisco J. Uribe}
\address{                    
Corresponding author \\
Department of Physics                 \\
Universidad Aut\'onoma Metropolitana  \\
M\'exico City, M\'exico 09340             \\
}

\author{
Wm. G. Hoover and Carol G. Hoover}      
\address{
Ruby Valley Research Institute         \\
Highway Contract 60, Box 601           \\
Ruby Valley, Nevada 89833              \\
}

\date{\today}
\begin{abstract}

We apply Maxwell and Cattaneo's relaxation approaches to the analysis of
strong shockwaves in a two-dimensional viscous heat-conducting fluid.
Good agreement results for reasonable values of Maxwell's relaxation times.
Instability results if the viscous relaxation time is too large.  These
relaxation terms have negligible effects on slower-paced subsonic
problems, as is shown here for two-roll and four-roll Rayleigh-B\'enard flow.

\end{abstract}

\vspace{0.1cm}

\maketitle

\section{Introduction}

In 1867 James Clerk Maxwell \cite{b1} noted that an initial shear stress in a
dilute gas,
(like air) when unsupported by an underlying shear motion, will decay
with a relaxation time $\tau = (\eta/P)$ (about 200 picoseconds for air),
where $\eta$ is the shear viscosity and $P$ the pressure. His governing
relaxation equation for the shear stress modifies Newton's $\sigma
= \eta \dot \epsilon$ to read
$$
\sigma + \tau \dot \sigma = \eta \dot \epsilon \ .
$$
Here $\sigma$ is the stress, $\eta$ the viscosity, and $\dot \epsilon$ 
the strain rate.  The superior dots represent comoving time
derivatives.

Nearly a century later Carlo Cattaneo \cite{b2} argued that Fourier's law
for heat conduction should be similarly modified, in order to avoid the
supersonic heat flow implied by a parabolic (diffusion equation)
transport law.  One could equally well argue that a heat flux, when
unsupported by a temperature gradient, would decay with a microscopic
relaxation time $\tau$ like Maxwell's.  Cattaneo's approach can be
written in a form like Maxwell's, but with a partial (fixed in space)
rather than a comoving time derivative:
$$
Q + \tau (\partial Q/\partial t) = -\kappa \nabla T \ .
$$
Cattaneo's rationale for using a partial time derivative rather than
one fixed in the material is unclear.  Here $Q$ is the heat flux, $T$ the
temperature, and $\kappa$ the heat conductivity.  With Cattaneo's
relaxation assumption, ``heat waves'' can propagate at about the speed of
sound \cite{b3}.  On physical grounds Maxwell's approach, with the comoving
time derivative, seems more ``realistic'' than Cattaneo's.  Cattaneo's
form for the relaxation time makes no contribution at all in stationary
steady-state problems such as the structure of a steady fluid shockwave.

Oddly enough, modern treatments of time delay \cite{b3,b4} often use
Cattaneo's partial-derivative formulation rather than Maxwell's
comoving time derivative.  The purpose of the present work is to elucidate
the usefulness of the relaxation concept and to explore its limits
in applications of fluid mechanics.  In the following Sections we
consider the relatively fast-paced steady shockwave problem as well
as the slower-paced steady convective Rayleigh-B\'enard flow.  A final
Section summarizes our findings.  For simplicity we use units in which 
the Boltzmann constant and atomic mass are both equal to one.  

\section{Strong Dense-Fluid Shockwaves}

The structure of strong shockwaves has long served as a testing ground for
continuum models like the Navier-Stokes-Fourier equations (here given for
a two-dimensional fluid with vanishing bulk viscosity, $\eta_V = 0$):
$$
\dot \rho = - \rho \nabla \cdot v \ ; \ \rho \dot v = -\nabla \cdot P \ ; \
\rho \dot e = - \nabla v : P - \nabla \cdot Q \ ; \
$$
$$
P = I[P_{\rm eq} + \eta \nabla \cdot v] - \eta [\nabla v + \nabla v^t] \ ;
\ Q = - \kappa \nabla T \ .
$$
The time derivatives, here as before indicated by the superior dot, are all
{\em comoving} derivatives, like Maxwell's, time rates of change in a
coordinate frame moving with the fluid velocity $v$.  Solving the three
differential equations for the density $\rho$, velocity $v$, and energy $e$
requires a knowledge of the pressure tensor $P$ and heat flux vector $Q$.
The simplest models are shown here, with two transport coefficients, the
Newtonian shear viscosity $\eta$ and the Fourier heat conductivity $\kappa$
defined in the usual way.  $I$ is the unit tensor, with
$I_{xx} = I_{yy} = 1$ and $ I_{xy} = I_{yx} = 0$.

Landau and Lifshitz' analytic solution of the shockwave structure for a gas
with constant transport coefficients and a shockwidth $\lambda$ provides a
useful initial condition for both macroscopic continuum and microscopic
molecular dynamics simulations \cite{b5}:
$$
\rho (x) = \frac{\rho_Ce^{-x/\lambda} + \rho_He^{+x/\lambda}}
                {      e^{-x/\lambda} +       e^{+x/\lambda}} \ 
\longrightarrow \ \{ \ v(x),P_{xx}(x),Q_x(x) \ \} \ .
$$
Their solution smoothly interpolates the density between cold fluid, with
density $\rho _C$, and hot fluid, with $\rho _H$.

Molecular dynamics shockwave simulations \cite{b6}-\cite{b15}
have been carried out in
the two different ways shown in Figure 1: (1) by following the two moving
waves generated by the inelastic collision of two blocks of material; (2) by
studying the single stationary wave formed with two boundary ``treadmills'' --
on the left boundary cold fluid is introduced at the ``shock
speed'' $v_s$ while at the right boundary hot fluid is extracted at the
slower speed $v_s-v_p$, where $v_p$ is the ``particle speed''.  In either case, in a coordinate frame centered on
the shockwave the mass, momentum, and energy fluxes are all constant:
$$
\{ \rho v,\ P_{xx} + \rho v^2, \ \rho v[e + (P_{xx}/\rho) + (v^2/2)] + Q_x \} \ 
{\rm constant \ for \ all \ } x \ .
$$
For ``weak'' shocks the Navier-Stokes-Fourier description is ``good'' \cite{b16}.
For ``stronger'' shocks (twofold compression) several contradictions to this
simple description arise
\cite{b8}-\cite{b15}.
To illustrate
these points typical mechanical and thermal shockwave profiles are shown in
Figure 2.

First, the local
longitudinal and transverse temperatures differ, often by more than a factor
of two (see Figure 2).  Second, as is also shown in Figure 2, the shear
stress ($P_{yy} - P_{xx}$)/2 and the heat flux $Q_x$ both lag behind the
velocity gradient $(dv_x/dx)$ and the temperature gradients ($dT_{xx}/dx$)
and ($dT_{yy}/dx$), suggesting the presence of Maxwell-type relaxation
times \cite{b12}-\cite{b15}.  Third, the fact that temperature is so very
anisotropic makes it necessary to consider separate $xx$ and $yy$
contributions to the heat flux \cite{b8}-\cite{b15}:
$$
Q_x = -\kappa_{xx} \, \nabla_x T_{xx} -\kappa_{yy} \nabla_x T_{yy} \ .
$$
Fourth, the same anisotropicity also suggests including asymmetric divisions
of the work and heat contributions (indicated by $\supset$) to the thermal energy
change:
$$
(\rho C_V/2)\, \dot T_{xx} \supset [-\alpha \nabla v : P_{\rm Thermal}  -\beta 
\nabla \cdot Q] \ ;
$$
$$
(\rho C_V/2) \, \dot T_{yy} 
\supset [-(1-\alpha) \nabla v : P_{\rm Thermal}  -(1-\beta) 
\nabla \cdot Q] \ .
$$
Here $C_V$ is the heat capacity per unit mass.
Fifth, a mechanism for the decay of temperature anisotropy must also
be included:
$$
[\dot T_{xx} - \dot T_{yy}] \supset 2[T_{yy} - T_{xx}]/\tau \ .
$$ 
Last, the molecular dynamics results imply that a bulk viscosity $\eta_V$,
approximately equal to the shear viscosity, must be included \cite{b12}.
Though a continuum model incorporating all of these ideas is necessarily
relatively complex, a successful implementation of all six of these additions
to the Navier-Stokes-Fourier model is described in References 
\cite{b11}, \cite{b13}, and \cite{b14}.

In those works all of the continuum field variables were derived from 
molecular dynamics simulations using a short-ranged repulsive pair potential,
$$
\phi(r<1) = (10/\pi)(1-r)^3 \ .
$$
The prefactor $(10/\pi)$ was chosen to give a potential energy integral
of unity for a random particle distribution at unit density:
$$
\int_0^1 2\pi r \phi(r) \equiv 1 \ .
$$
The initial zero-pressure zero-temperature state was compressed twofold to
obtain a hot dense fluid state.  Lucy's normalized weighting
function \cite{b17,b18} was used to compute spatial averages of the various
field variables:
$$
w(r<h) = (5/\pi h^2)[1-(r/h)]^3[1+3(r/h)] \rightarrow \int_0^h 2\pi r w(r)
\equiv 1 \ .
$$
The {\em smooth-particle} average of the particle quantity $f_j$ is given
by a weighted sum,
$$
\langle \rho (r) f(r) \rangle = \sum _jm_jf_jw(r-r_j) \ ; \ 
\rho(r) \equiv \sum_jm_jw(r-r_j) \ .
$$
This smooth-particle definition has two advantages: (1) all of the field
variables defined in this way have two continuous space derivatives; (2)
the continuity equation (with $f_j$ equal to the particle velocity $v_j$)
is satisfied exactly:
$$
\{ \ \rho = \sum_jm_jw(r-r_j) \ ; \ \rho v = \sum_jm_jv_jw(r-r_j) \ \}
\ \longrightarrow \ \dot \rho \equiv - \rho \nabla \cdot v \ .
$$
Here $\rho$ and $\rho v$ are defined {\em everywhere} in this way, not just at
the particle locations.  The range $h$ of the ``weighting function'' 
$w(r<h)$ is typically chosen so that about 20 particles contribute to
field-point averages. With this approach the microscopic pressure tensor
and heat flux vector at any point in space are expressed in terms of 
nearby individual particle contributions to these nonequilibrium
fluxes \cite{b19,b20}.

To appreciate the effect of the various modifications of the
Navier-Stokes-Fourier model we next study the stability of solutions using
a continuum model which is a rough representative of the molecular dynamics
results \cite{b10,b11,b12}.

\section{Stability Studies with an Idealized Gr\"uneisen Model}

For stability studies we choose an equilibrium equation of state based
on Gr\"uneisen's separation of the energy and pressure into cold and thermal
parts:
$$
P_{\rm eq} = \rho e = (\rho^2/2) + 2\rho T \ ; \ e = (\rho/2) + 2T \ .
$$
A shockwave satisfying all the conservation laws results when a cold fluid
is compressed to twice its initial density by a shockwave moving toward
that fluid at twice the particle velocity ($v_s = 2v_p = 2$).  In this
case the constant mass, momentum, and energy fluxes are respectively
$$
\{ \ \rho v = 2 \ ; \ P_{xx} + \rho v^2 = (9/2) \ ; \
\rho v[e + (P_{xx}/\rho) + (v^2/2)] + Q_x = 6 \ \} \ .
$$
The various hydrodynamic variables then cover the following ranges within
the shockwave: 
$$
[ \ 2 > v(x) > 1 \ ] \ ; \ [\ 1 < \rho(x) < 2 \ ] \ ;
\ [ \ (1/2) < e(x) < (5/4) \ ] \ ;
$$
$$
[ \ (1/2) < P_{\rm eq} < (5/2) \ ] \ ; \
[ \ 0 < T_{\rm eq} < (1/8) \ ] \ . 
$$
(Note that $T_{xx}$ can exceed the ``hot'' value of (1/8) within the
shockwave.)
The details of the shockwave structure depend upon the nonequilibrium
constitutive relations for the shear stress and the heat flux.  Next, we
summarize two separate situations, (1) vanishing conductivity with a scalar
temperature; (2) tensor conductivity, with separate longitudinal and
transverse temperatures, with different contributions from work and heat.
Both these models lead to the conclusion that the mechanical
relaxation time cannot be too large.  By contrast, the thermal relaxation
time can be either ``small'' or ``large''.

\subsection{Relaxation Without Heat Conduction}

The simplest case results when both heat conductivity and thermal anisotropy
are omitted.  Then the density and energy can both be eliminated from the
three flux equations,
$$
\rho v = 2 \ ; \ (\rho e) - \sigma + 2v = (9/2) \ ; \
2[e + e - (v/2)\sigma + (v^2/2)] = 6 \ ,
$$
giving the shear stress,
$$
\sigma = (P_{yy} - P_{xx})/2 = \rho e - P_{xx} \ , 
$$
as a function of velocity:
$$
\sigma = (3/v)(v-1)(v-2) < 0 \ .
$$
Evidently the viscous stress is everywhere negative (compressive).  If we
introduce Maxwell's idea of comoving stress relaxation,
$$
\sigma + \tau \dot \sigma = \sigma + \tau v (d\sigma /dx) =
\sigma + \tau v (d\sigma /dv)(dv/dx) =\eta(dv/dx) \ ,
$$
we find that the velocity gradient  $(dv/dx)$ diverges unless the ratio
$(\tau / \eta)$ is sufficiently small:
$$
\tau _{\sigma} < (\eta /3) \ .
$$
It is physically reasonable that too long a memory can lead to instability in
fast-paced complex flows like shockwaves.  On the other hand the relaxation
equation by itself, with a smooth strain increment localized near zero time
($t=0$),
$$
\sigma + \tau \dot \sigma = \frac{1}{[e^{-t} + e^{+t}]} \ ,
\vspace{0.5 cm}
$$
provides smooth solutions even for large $\tau $ \cite{b13,b14}.  The present
analytic shockwave limit on $\tau _\sigma < (\eta/3)$ is in full accord
with two kinds of numerical simulations. First, the stationary flux
equations can be solved for the temperature and stress fields, just
as was indicated above for the case of vanishing conductivity.  Second, it is
possible to solve the dynamical equations for 
$$
\{ \ (\partial \rho /\partial t),(\partial v /\partial t),
(\partial e /\partial t) \ \} \ {\rm or} \ 
\{ \ (\partial \rho /\partial x),(\partial v /\partial x),
(\partial e /\partial x) \ \} \ 
$$
starting with the Landau-Lifshitz profile.
 The two methods agree.  They
show that the stress relaxation time in shockwaves must be sufficiently small,
$\tau _\sigma < (\eta/3)$ for stability.

We next extend the thermal constitutive model to include tensor temperature
with anisotropic heat conduction. We also include separate relaxation
times for the longitudinal and transverse heat fluxes, and the separation of
work and heat into longitudinal and transverse parts \cite{b11,b13}.

\subsection{Lack of Relaxation Without Viscosity}

Viscosity, as opposed to heat conduction, is essential to the shock process.
To appreciate this need, consider the conservation equations for our simple
model {\em without} viscosity and with the heat conductivity equal to unity:
$$
e= (\rho /2) + 2T \ ; \ \rho v = 2 \ ; \ \rho e + \rho v^2 = (9/2) \ ; \
\rho v[2e + (v^2/2)] - (dT/dx) = 6 \ .
$$
According to the first three equations the temperature has its maximum value
of $(T_{\rm max} = 0.238 > T_{\rm hot} = 0.125)$ within the shock:
$$
\{ \rho, v, T \} = \{ 1.4436, 1.38545, 0.23800 \} \ {\rm for} \ T = T_{\rm max} \ .
$$
But the fourth (energy-flux) equation gives $(dT/dx) = 0.7106$ for that
thermodynamic state, contradicting the presence of a maximum.  Thus this
model, lacking viscosity, cannot sustain a stationary shockwave.

Exactly this same conclusion follows also for the inviscid ideal gas, with
twofold compression from unit density, pressure, and temperature, with
$v_s = \sqrt{8}$ and the wholly thermal pressure $P = \rho e = \rho T$.
Because heat conductivity in the absence of viscosity is not enough to
provide a shockwave, the relaxation effects are quite different for
conductivity and viscosity, as we show next.

\subsection{Relaxation with Tensor Temperature, Apportioned Work and Heat}

The analysis becomes more complicated when heat flow is included, along with
relaxation and separated contributions of the heat and work to the
longitudinal and transverse temperatures.  Here the heat flux evolves
following the tensor relaxation equation:
$$
Q_x + \tau _Q\dot Q_x = -\kappa_{xx}(dT_{xx}/dx) - \kappa_{yy}(dT_{yy}/dx) \ .
$$
The divergence of the heat flux provides net heating and is apportioned
between the longitudinal and transverse temperatures:
$$
\rho \dot T_{xx} \supset -\beta(dQ_x/dx) \ ; \
\rho \dot T_{yy} \supset –(1 - \beta)(dQ_x/dx)  \ .
$$
The contributions of the heat flux divergence $\nabla \cdot Q$ to heating are
indicated by the inclusion symbol, ``$\supset$''.  We include also
an analogous separation of the thermodynamic work into longitudinal and
transverse parts:
$$
\rho \dot T_{xx} \supset -\alpha P_{\rm Thermal}:\nabla v \ ; \
\rho \dot T_{yy} \supset –(1 - \alpha)P_{\rm Thermal}:\nabla v \ .
$$
Finally, the two temperatures necessarily relax toward one another:
$$
\dot T_{xx} \supset (T_{yy} - T_{xx})/\tau _Q \ ; \
\dot T_{yy} \supset (T_{xx} - T_{yy})/\tau _Q \ .
$$
For simplicity we choose the two thermal relaxation times [ for the
heat flux $Q$ and the temperature anisotropicity $(T_{xx}-T_{yy})$ ] to have
a common value, $\tau _Q$.  For illustrative purposes we emphasize the
difference between the two temperatures by choosing the apportionment
parameters $\alpha$ and $\beta$ both equal to unity, so that both
the work and the heat provide longitudinal heating, with the transverse
temperature lagging behind.

Then straightforward (at least for a computer) algebra provides solutions
of the shockwave problem and reveals not one, but {\em two} restrictions
on $\tau _Q$.  For stable solutions to exist we found in this way
that the thermal relaxation time must be either sufficiently small or
sufficiently large.  Setting the distance scale of the shockwave with the
constant transport coefficients 
$$
\eta = 2\kappa_{xx} = 2\kappa_{yy} = 1 \ ,
$$
computer algebra gives the following restrictions on the relaxation times:
$$
0 < \tau _\eta < (1/3) \ ; \
 \tau _Q < (1/8) \ {\rm or} \ \tau _Q > (1/4) \ .
$$

Figures 3 and 4 show typical continuum profiles using these constitutive
relations.  The continuum profiles were generated in two quite different
ways: (1) solving the time-dependent equations for
$\{ \rho, v, e ,\sigma, Q \}$ starting with the Landau-Lifshitz
approximation; (2) solving the stationary flow
equations for the mass, momentum, and energy fluxes using a computer algebra
program (we used ``Maple'').  The latter approach provides page-long formul\ae
\ for $(du/dx)$, $(dT_{xx}/dx)$, and $(dT_{yy}/dx)$ as well as numerical
solutions.  The stationary equations for the shockwave profile
have no solution if the relaxation time for the shear stress $\tau _\sigma $
is greater than $(\eta/3)$ or if the relaxation time for the heat flux lies
between $(\kappa/8)$ and $(\kappa/4)$.

To summarize, our findings for shockwaves establish that momentum-flux
relaxation has to be ``fast'' for stability.  Thermal relaxation can
either be likewise fast or quite slow, with a window of instability
separating these two regimes.  Where the thermal relaxation is slow
the shockwave structure is
dominated by viscosity rather than conductivity.

It is natural to speculate on the effect of relaxation in ordinary
hydrodynamic situations.  In order to see what consequences arise from 
these effects in subsonic fluid mechanics we next introduce delay into the
hydrodynamic equations describing a compressible, conducting, viscous
flow, the Rayleigh-B\'enard problem.

\section{Rayleigh-B\'enard Flow}

To investigate the stability of moderate flows to the presence of viscous
and thermal relaxation we revisit some finite-difference Rayleigh-B\'enard
simulations of two-roll, four-roll, and six-roll flows \cite{b21,b22}.  The
simulations picture a viscous conducting fluid, heated from below in the
presence of a vertical gravitational field.  Sufficiently strong heating
causes a transition from static heat conduction to one of a number of
nonequilibrium steady states with stationary convection rolls.  Stationary
and transient sample flows are shown in Figures 5 and 6.

For the Rayleigh-B\'enard model we study here (equal kinematic viscosity
and thermal diffusivity) the transition from static Fourier conduction to
two-roll convection occurs near a Rayleigh number $R$ of 1750:
$$
R = g(\partial \ln V/\partial T)_PH^3\Delta T/(\nu D_T) = H^2/(\nu D_T) \ .
$$
The fluid is confined to a rectangular box, periodic on the sides, with the
gravitational constant $g = (1/H)$ chosen to give constant density in the
nonconvecting case.  $H$ is the height of the cell, equal here to half the
width.  $\Delta T$ is the difference between the hot temperature at the base
($T_H = 1.5$) and the cold temperature at the top of the cell ($T_C = 0.5$).
$\nu $ and $D_T$ (chosen equal, for convenience) are the kinematic viscosity
and thermal diffusivity (both with units of $[length^2/time]$).  For
simplicity we choose all values of the relaxation times equal and do not
distinguish between the longitudinal and transverse temperatures,
$T_{xx} = T_{yy}$.  Our model continuum fluid obeys the ideal gas equation
of state:
$$
P_{\rm eq} = \rho T = \rho e \ ; \ \eta_V = 0 \ ;
\ \eta = 2\kappa_{xx} = 2\kappa_{yy} = 1 \ .
$$

Numerical results for this model are given as a function of Rayleigh number
in References \cite{b21} and \cite{b22}.  
Simulations with the various relaxation times all
equal to 0.1 reproduced this earlier work perfectly.  As an example, the
two-roll problem of References \cite{b21} and \cite{b22}, 
with a Rayleigh number of 40,000
gives per-cell kinetic energies of $(K_x/N) + (K_y/N) = 0.00373 + 0.00357$.
We carried out many special cases with a Rayleigh Number of 40,000, which
produces stationary steady states.  Whether two-roll or four-roll solutions
are obtained is sensitive to the initial conditions \cite{b22}.  We began
with a very weak two-roll velocity field as the initial condition in an
$H \times W$ box with the coordinate origin at its center:
$$
v_x \propto \sin(2\pi x/W)\sin(2\pi y/H) \ ; \
v_y \propto \cos(2\pi x/W)\cos( \pi y/H) \ . 
$$
We found solutions for $\tau = 5\eta = 5\kappa$ and $\tau = 10\eta = 10\kappa$
but instability when $\tau$ was doubled again to $20\eta = 20\kappa$.
These additions of relaxation to the Navier-Stokes-Fourier equations lowered the
horizontal kinetic energy and raised the vertical, with both effects on
the order of parts per thousand.  Thus relaxation in subsonic flows has only
relatively small effects in the regime of stable solutions.

\section{Conclusions}

Molecular dynamics simulations have established the facts that delay times
on the order of a collision time, as envisioned by Maxwell, affect shockwave
structure in a substantial way.  Cattaneo's approach, with partial time
derivatives, has no effect on shockwave structure.  Shockwaves are dominated
by viscosity, so that stress relaxation must be relatively rapid.  Thermal
relaxation, important for chemical relaxation, can be either fast or slow.

In ordinary subsonic fluid mechanics the effects of time delays are relatively
small.  As a result, thermal anisotropicity is ordinarily ignored in continuum
mechanics.  It is a
substantial effect in shocks, with repercussions for chemical reaction rates.
In our continuum simulations we have assumed relaxation equations 
with comoving time derivatives,
$$
\sigma + \tau_\eta \dot \sigma =  \eta \dot \epsilon \ , \
Q + \tau_Q \dot Q =  -\kappa \nabla T \ ,
$$
rather than partial derivatives.  If $\dot \sigma$ were replaced with
$(\partial \sigma /\partial t)$ there would be no relaxation at all in a
stationary problem like the shockwave and Rayleigh-B\'enard problems
studied here.

The Maxwellian relaxation times cause no trouble solving conventional moderate
flow problems like Rayleigh-B\'enard convection.  The problem areas suggested
by this work include (1) formulating optimum choices for locally-averaged
hydrodynamic variables with the general goal of maximizing the accuracy of
macroscopic descriptions of microscopic results and (2) developing 
theoretical models for the estimation of the relaxation parameters measured
in the dynamical simulations.

A logical approach to problem (1) above would use ``entropy production" as a
tool \cite{curtiss}.  
In the Rayleigh-B\'enard problem entropy production is proportional to the
squares of the nonequilibrium fluxes, $\sigma ^2$ and $Q^2$.  If these are
computed locally, with a weight function $w(r<h)$ then $h$ can be chosen such
that the internal entropy production matches the boundary sources and sinks of
entropy.  Evidently too small/large an $h$ gives too large/small an entropy production,
so that $h$ can be chosen to be ``just right".  Problem (2) would have to begin with
some nonequilibrium simulations tailored to the direct measurement of delay and relaxation.

Finally, the presence of delay has some pedagogical importance.  Delay
in the results of time-reversible motion equations (molecular dynamics) 
breaks the time-symmetry which would otherwise lead to a logical
contradiction between time-reversible molecular dynamics and conventional
irreversible continuum mechanics \cite{b12}.

\section{Acknowledgments}

We thank David Sanders and Thomas Gilbert for organizing a 2011 Workshop,
``Chaotic and Transport Properties of Higher-Dimensional Dynamical Systems'',
in Cuernavaca (M\'exico), at which the authors were able to work together.  
We are
specially grateful to Brad Holian for his continuing interest in this work
and for supplying prepublication copies of Reference 15.

\section{Figure captions}

\begin{itemize}
\item{Figure 1.}
Two colliding fluid blocks generate symmetric shockwaves (velocities
$\pm v_p$) as the blocks, moving at $\pm[v_s - v_p]$ collide and come
to a stop (shown above). Two treadmill boundaries, one fast
(velocity $v_s$) and one slow (velocity $v_s - v_p$), maintain a
single stationary shockwave in the center of the system (shown below).

\item{Figure 2.}
Density, pressure, internal energy, temperature tensor, and heat flux in a
strong shockwave simulation using molecular dynamics.  In the cold unshocked
material the nearest-neighbor spacing is unity.  The hot shocked fluid has a
density exactly twice that of the cold material.
Figure based on data described in reference \cite{b14}.

\item{Figure 3.}
Solution of the continuum model for twofold compression with the
Gr\"uneisen equation of state using $\tau_\eta = (1/10)$ and 
$\tau_Q = \tau_R = 1$.  The mass, momentum, and energy fluxes are
$\{2, (9/2), 6\}$  Compare with Figure 4 noting particularly the
differences between $T_{xx}$ and $T_{yy}$.

\item{Figure 4.}
Solution of the continuum model for twofold compression with the
Gr\"uneisen equation of state using $\tau_\eta = \tau_Q =
\tau_R = (1/10)$.  The mass, momentum, and energy fluxes are
$\{2, (9/2), 6\}$  Compare with Figure 3.

\item{Figure 5.}
Transient flow field for the Rayleigh-B\'enard problem at time = 1000.
The initial state was two weakly-rotating rolls.  The viscous,
heat-conducting, compressible fluid is  heated at the bottom and cooled
at the top with a gravitational field directed downward.  The vertical
boundaries at the sides are periodic.  The number of computational
cells shown here is $80 \times 40 = 3200$.  The transport coefficients,
$\eta = \kappa = (1/5)$ were selected to give a Rayleigh number of
40,000.  The relaxation times were set equal to unity: $\tau_\eta =
\tau_Q \equiv 1$.

\item{Figure 6.}
A fully-converged four-roll structure evolved from the flow field shown
in Figure 5.  Here the time is 10,000.  The Rayleigh number is 40,000
and the viscosity and heat conductivity, $\eta = \kappa = (1/5)$,  have
equal relaxation times, $\tau_\eta = \tau_Q \equiv 1$.  The final
kinetic energy is $(K_x/N) + (K_y/N) = 0.001144 + 0.004133 = 0.005277$.
The number of computational cells is $N = 80 \times 40$.

\end{itemize}

\vfill \eject

\begin{figure}[t]
\vspace{1 cm}
\includegraphics[height=12cm,width=8cm,angle=-90]{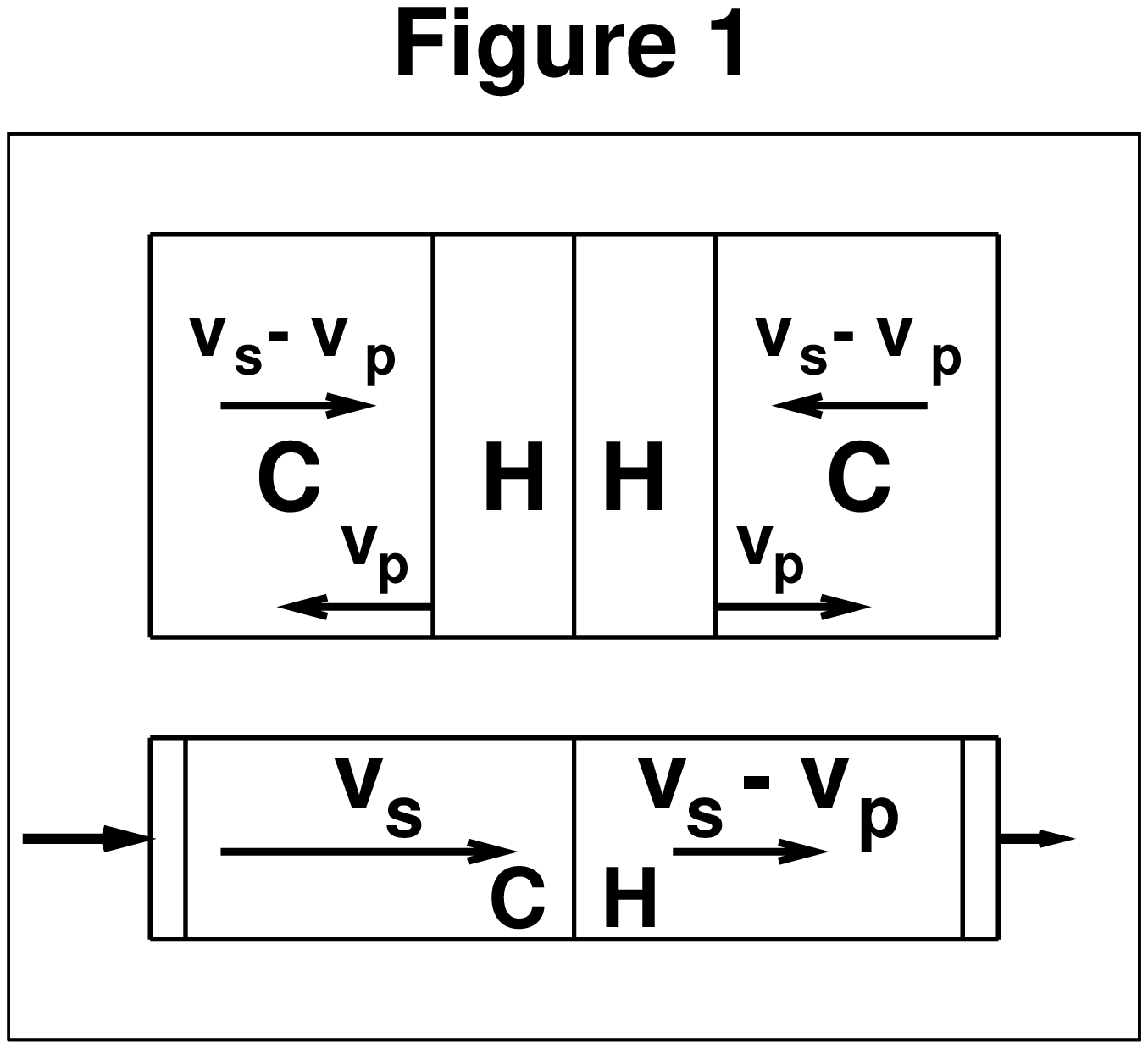}
\vspace{8 cm}
\end{figure}

\vfill\eject

\begin{figure}[t]
\vspace{1 cm}
\includegraphics[height=14cm,width=10cm,angle=-90]{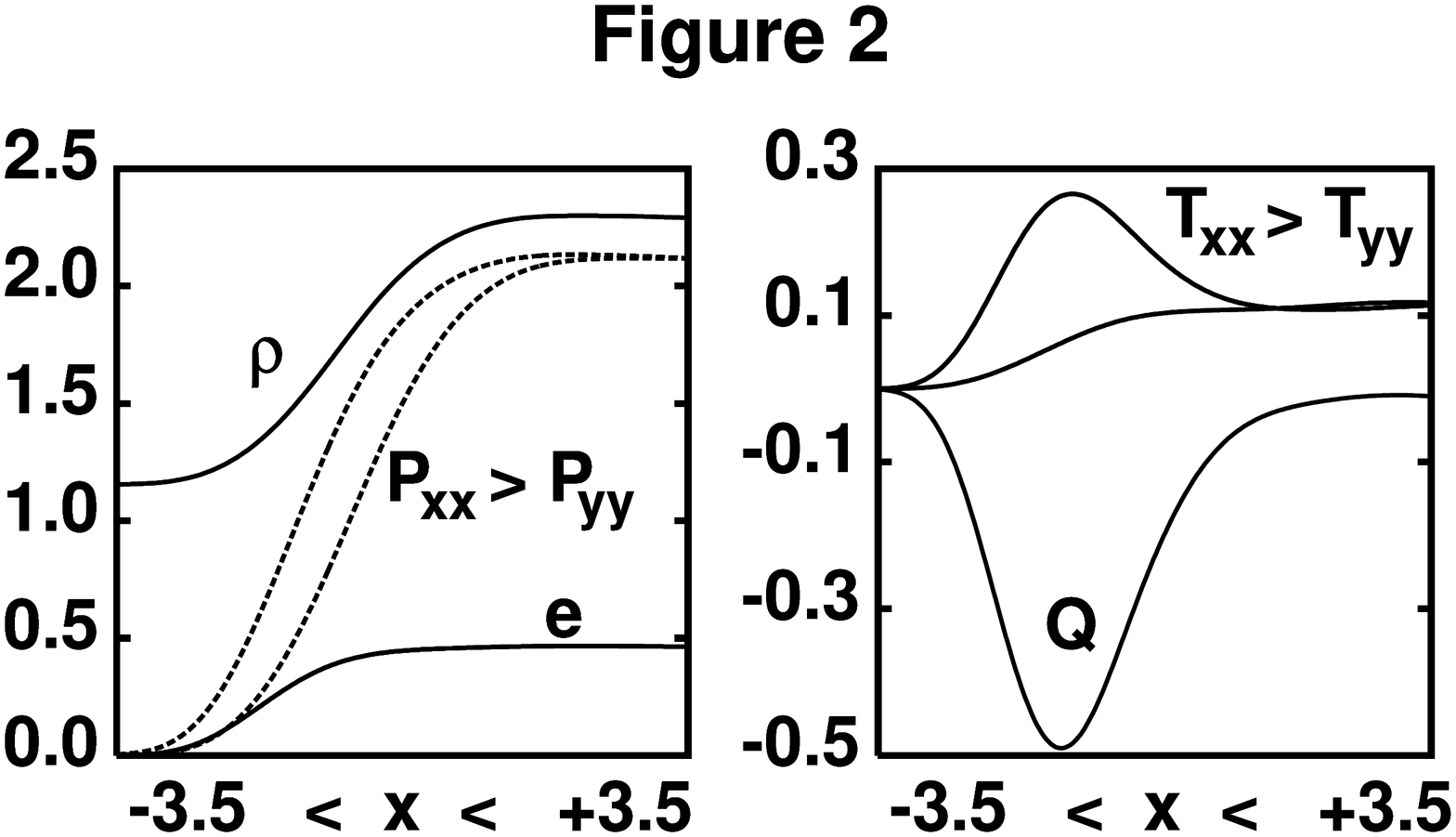}
\vspace{8 cm}
\end{figure}

\vfill\eject
\begin{figure}
\vspace{2 cm}
\includegraphics[height=14cm,width=10cm,angle=-90]{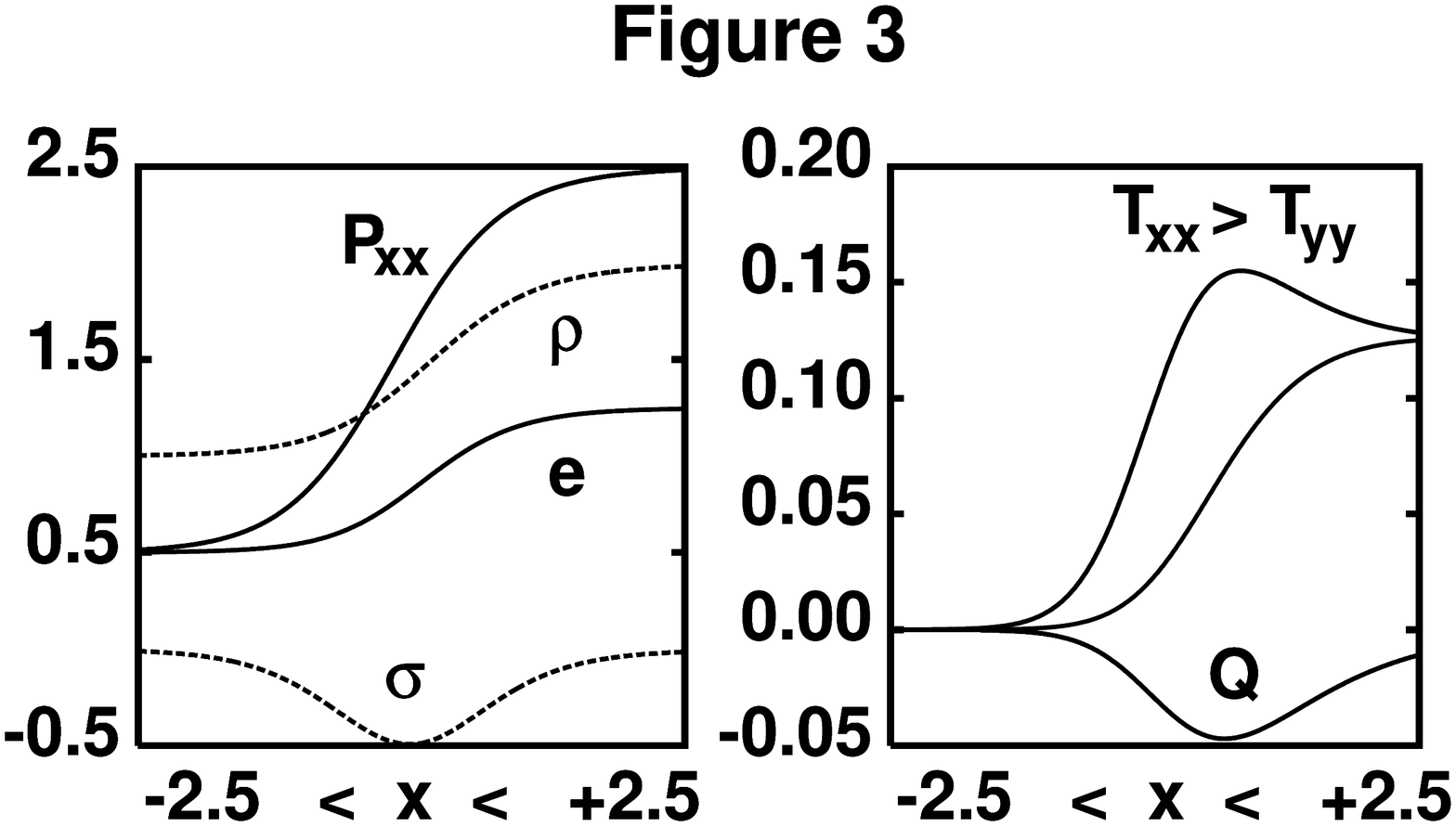}
\vspace{8 cm}
\end{figure}

\vfill\eject
\begin{figure}
\vspace{1 cm}
\includegraphics[height=14cm,width=10cm,angle=-90]{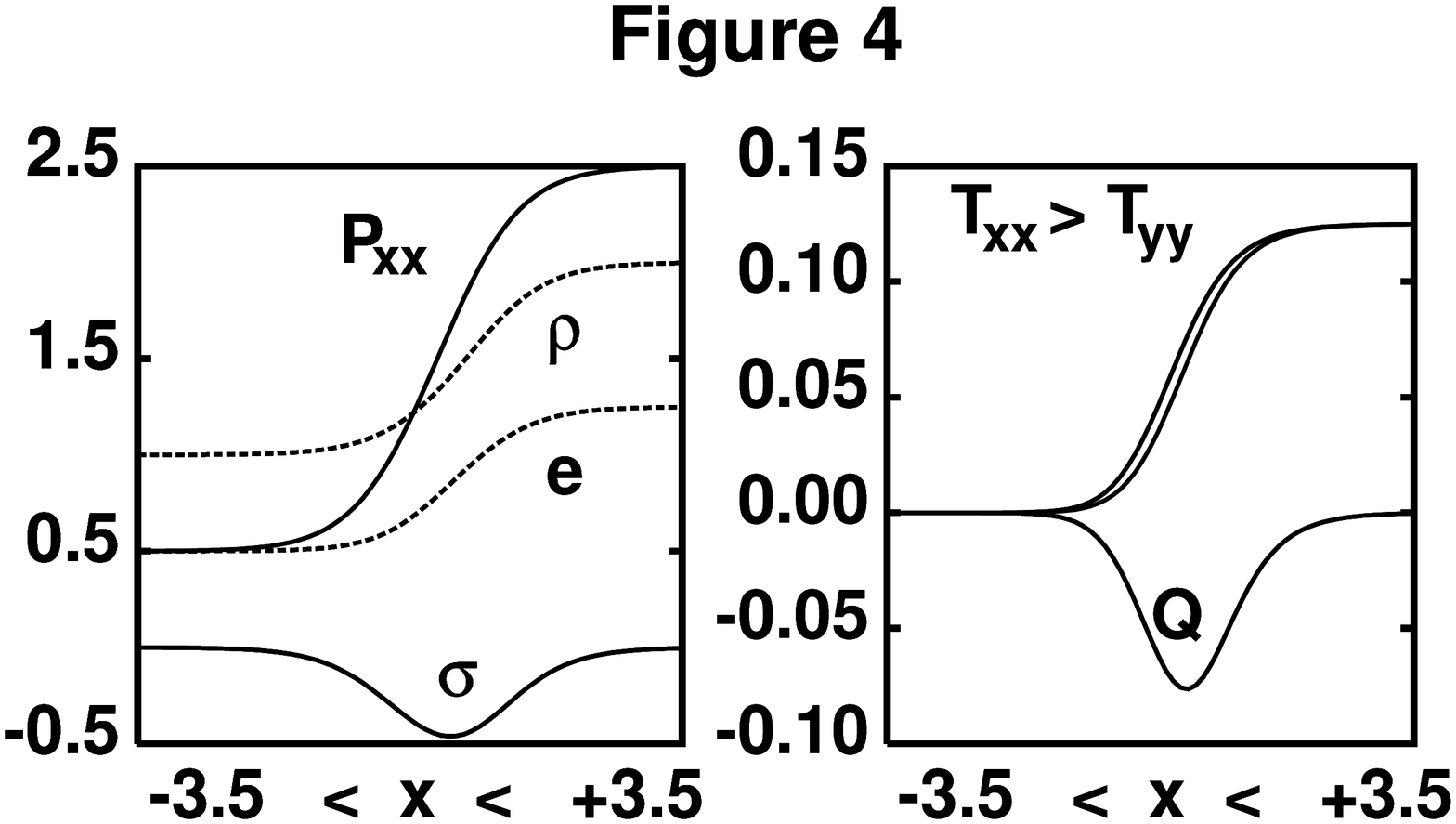}
\vspace{8 cm}
\end{figure}

\vfill\eject
\begin{figure}
\vspace{1 cm}
\includegraphics[height=14cm,width=10cm,angle=-90]{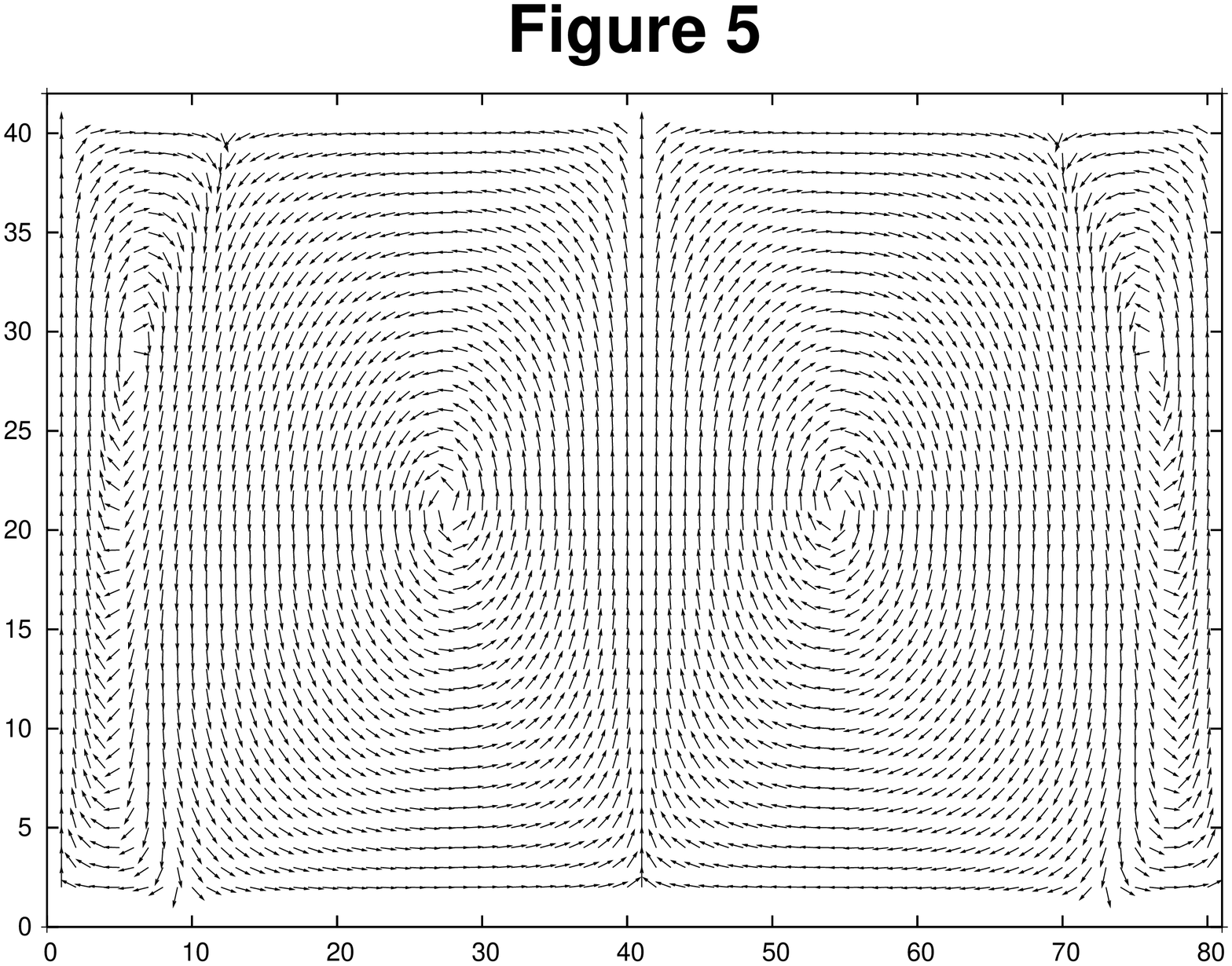}
\vspace{8 cm}
\end{figure}

\vfill\eject
\begin{figure}
\vspace{1 cm}
\includegraphics[height=14cm,width=10cm,angle=-90]{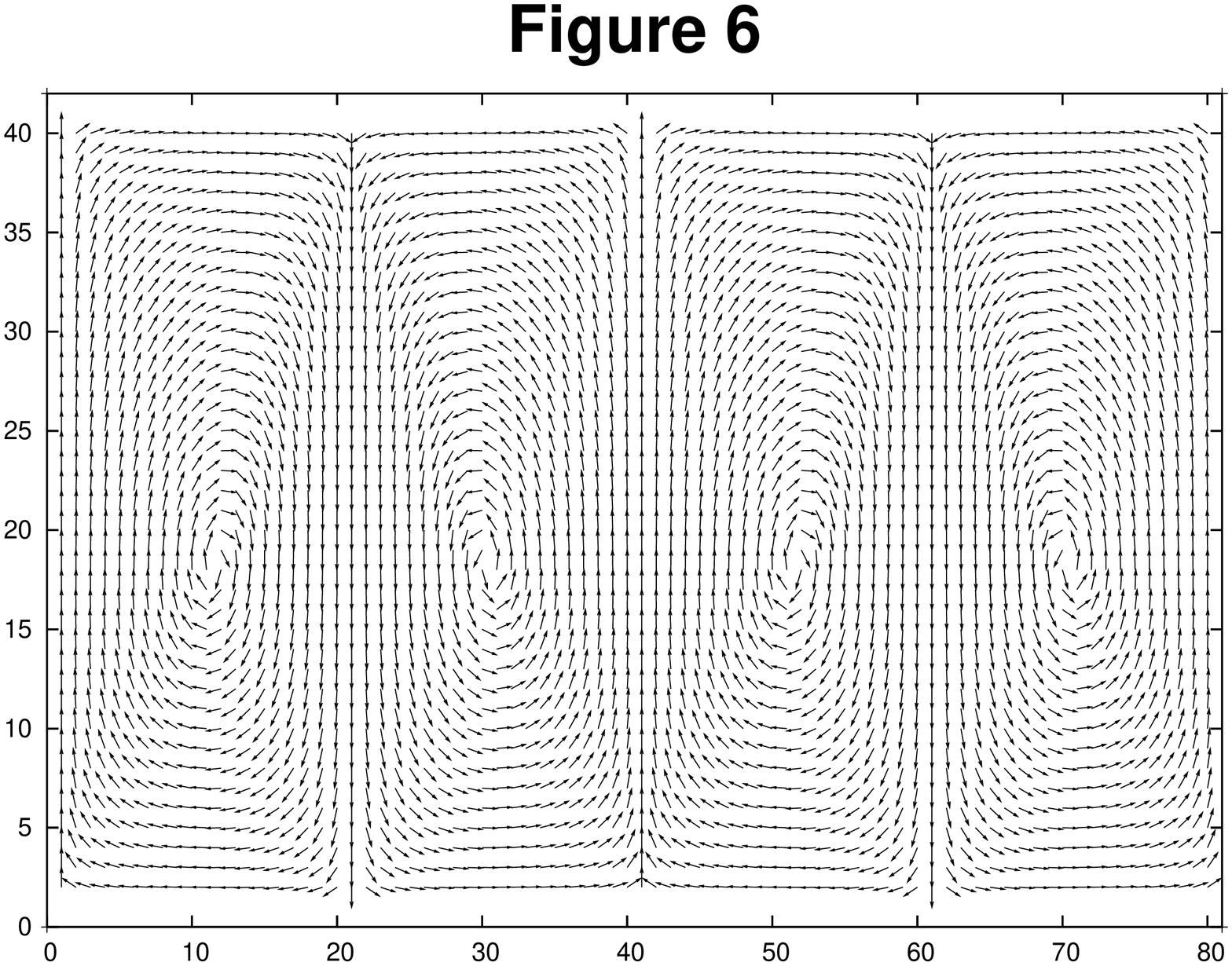}
\vspace{8 cm}
\end{figure}

\vfill\eject 


\end{document}